\documentclass[a4paper, conference]{IEEEtran}
\usepackage[utf8]{inputenc} 
\usepackage[T1]{fontenc}
\usepackage{url}
\usepackage{ifthen}
\usepackage{amsfonts, amsthm, amssymb, dsfont}
\usepackage[cmex10]{amsmath} 
\usepackage{url}
\usepackage{hyperref}

\usepackage{cite}

\usepackage[a4paper, top=1.1in, bottom=1.1in, left=1.5cm, right=1.5cm]{geometry} 



\IEEEoverridecommandlockouts

\newtheorem{theorem}{Theorem}
\newtheorem{lemma}{Lemma}
\newtheorem{corollary}{Corollary}
\newtheorem{remark}{Remark}

\makeatletter
\newcommand{\vast}{\bBigg@{3}}
\newcommand{\Vast}{\bBigg@{4}}
\makeatother

\newcommand{\bs}[1]{\boldsymbol{#1}}
\newcommand{\mcl}[1]{\mathcal{#1}}
\newcommand{\msf}[1]{\mathsf{#1}}

\newcommand{\I}{\mathrm{I}}

\begin{document}
\title{Class-Based Expurgation Attains Csisz\'ar's Expurgated Source-Channel Exponent}
%\title{Achieving Csisz\'ar's Expurgated Source-Channel Coding Exponent Using Two Classes}
%TODO: Class-Based Expurgated Exponent Coding in Source Channel Coding
%Class-Based Expurgation for Achieving Csisz\'ar’s Source–Channel Exponent

\author{%
  \IEEEauthorblockN{Seyed AmirPouya Moeini}
  \IEEEauthorblockA{University of Cambridge\\
    \texttt{sam297@cam.ac.uk}}
  \and
  \IEEEauthorblockN{Albert Guill\'en i F\`abregas}
  \IEEEauthorblockA{University of Cambridge\\
  Universitat Polit\`ecnica de Catalunya\\
    \texttt{guillen@ieee.org}}
  \thanks{This work was supported in part by the European Research Council under Grants 725411 and 101142747, and in part by the Spanish Ministry of Economy and Competitiveness under Grant PID2020-116683GB-C22.}
}

\maketitle

\begin{abstract}
This paper studies expurgated error exponents for joint source-channel coding for discrete memoryless sources and channels. We consider a partition of the source messages into classes, where the codeword distributions depend on the class. We show that two carefully chosen classes suffice to achieve Csisz\'ar's expurgated exponent.
\end{abstract}

%%%%%%%%%%%%%%%%%%%%%%%%%%%%%%%%%%%%%%%%%%%%%%%%%%%%%%%%%%%%%%%%%%%%%%%%%%%%%%%%%%%%%%%%%%%%%
%%%%%%%%%%%%%%%%%%%%%%%%%%%%%%%%%%%%%%%%%%%%%%%%%%%%%%%%%%%%%%%%%%%%%%%%%%%%%%%%%%%%%%%%%%%%%
\section{Introduction}
We study the transmission of non-equiprobable messages from a discrete memoryless source (DMS) with distribution $P^k(\bs{v}) = \prod_{i=1}^k P_V(v_i)$, where $\bs{v} = (v_1, ..., v_k) \in \mcl{V}^k$ is the source message, and $\mcl{V}$ is a finite discrete alphabet. 
The channel is a discrete memoryless channel (DMC) given by $W^n(\bs{y}|\bs{x}) = \prod_{i=1}^n W(y_i | x_i)$, $\bs{x}=(x_1, \ldots, x_n) \in \mcl{X}^n$ and $\bs{y}=(y_1, \ldots, y_n) \in \mcl{Y}^n$,
where $\mcl{X}$ and $\mcl{Y}$ are discrete
alphabets with cardinalities $|\mcl{X}|$ and $|\mcl{Y}|$, respectively.
An encoder maps the length-$k$ source message $\bs{v}$ to a length-$n$ codeword $\bs{x}_{\bs{v}}$, which is then transmitted over the channel. 
We refer to $t \triangleq k / n$ as the transmission rate. Based on the channel output $\bs{y}$ the decoder guesses which source message was transmitted.

We say that an error exponent $E>0$ is achievable if there exists a sequence of codes of length $n$ such that the error probability satisfies
\begin{align}
	p_e \leq e^{-n E+o(n)},
\end{align}
where $\lim _{n \rightarrow \infty} o(n) / n=0$. 
Most achievability results are obtained using random coding, followed by analyzing the average probability of error.
% or inducing certain properties to derive error exponents, such as the expurgated exponent.
It has been shown that joint source-channel coding (JSCC) can achieve lower error probabilities than separate source and channel coding~\cite{csiszar1980}.
 In Csisz\'ar's construction, codewords are randomly selected from a set of sequences whose composition depends on the source message composition. 
% This construction achieves the best exponent for that setting.
Later, Zhong et al.\cite{1614076} studied the random coding error exponent in JSCC in more detail and proved that Csisz\'ar's exponent corresponds to the concave hull of the $E_0$ function in the dual domain. 
A more general message set partitioning method was studied in \cite{6803047}, where it was shown that two carefully chosen classes are enough to achieve Csisz\'ar's random coding exponent.

Expurgation has received less attention than random coding bounds in JSCC. 
%In~\cite{1056585}, Csisz\'ar derived two expurgated exponents, which were later recovered by Scarlett~\cite{8006660} using Gallager's classical expurgation method~\cite[Sec.~5.7]{gallager}.
In~\cite{1056585}, Csisz\'ar derived two expurgated exponents, which were later recovered by Scarlett~\cite{8006660} using a new method based on source-type duplication and type-by-type expurgation.

Motivated by the preceding discussion, we study expurgation under a partition of the source messages. 
We show that, as in the random-coding case, two carefully chosen classes are sufficient to achieve Csisz\'ar's second expurgated exponent \cite{1056585}.

%Due to space limitations, technical details and complete proofs are omitted and are provided in the full version~\cite{moeini2025classbasedexpurgationattainscsiszars}.

%\subsection*{Notations}
\emph{Notation:}
In this paper, scalar random variables are denoted by uppercase letters, their realizations by lowercase letters, and their alphabets by calligraphic letters. Random vectors are written in boldface.
For two positive sequences $f_n$ and $g_n$, we write $f_n \doteq g_n$ if $\lim _{n \rightarrow \infty} \frac{1}{n} \log \frac{f_n}{g_n}=0$, and we write $f_n \, \dot{\leq}\,\, g_n$ if $\lim \sup _{n \rightarrow \infty} \frac{1}{n} \log \frac{f_n}{g_n} \leq 0$.

The type of a sequence $\bs{x}=\left(x_1, ..., x_n\right) \in \mcl{X}^n$ is its empirical distribution, defined by
$\hat{P}_{\bs{x}}(x) \triangleq \frac{1}{n} \sum_{i=1}^n \mathds{1}\left\{{x}_i=x\right\} .$
The set of all probability distributions on an alphabet $\mcl{X}$ is denoted by $\mcl{P}(\mcl{X})$, while $\mcl{P}_n(\mcl{X})$ represents the set of empirical distributions for vectors in $\mcl{X}^n$.
For $P_X \in \mcl{P}_n(\mcl{X})$, the type class $\mcl{T}^n(P_X)$ consists of all sequences in $\mcl{X}^n$ with type $P_X$.
%It is shown in \cite[Lemma~2.3]{csiszar2011information} that $|\mcl{T}^n(P_X)| \doteq \exp(nH(P_X))$.

%We assume that the source set $\mcl{V}^k$ contains $N_k$ source type classes, which grow polynomially with $k$~\cite[Lemma~2.2]{csiszar2011information}. 
The number of source type classes in the set $\mcl{V}^k$ is denoted by $N_k$, which grows polynomially with $k$~\cite[Lemma~2.2]{csiszar2011information}. 
Throughout, we use the indices $i$ and $j$ to refer to source types, and $c$ to indicate the class to which a source message belongs under a given partitioning. 
The total number of classes in the partitioning is denoted by $m$.
Moreover, for a source type $P_i$, we define $R_i \triangleq tH(P_i)$.

%%%%%%%%%%%%%%%%%%%%%%%%%%%%%%%%%%%%%%%%%%%%%%%%%%%%%%%%%%%%%%%%%%%%%%%%%%%%%%%%%%%%%%%%%%%%%
%%%%%%%%%%%%%%%%%%%%%%%%%%%%%%%%%%%%%%%%%%%%%%%%%%%%%%%%%%%%%%%%%%%%%%%%%%%%%%%%%%%%%%%%%%%%%

\section{Csisz\'ar's Expurgated Error Exponents}
In this section, we introduce definitions and summarize existing expurgated results in JSCC that are relevant to this work.
The source reliability function \cite[Ch.~9]{csiszar2011information} is given by
\begin{align}
	e(R, P_V) &\triangleq \min_{Q: H(Q) \geq R} D(Q \| P_V),
\end{align}
or, in the dual domain \cite{Jel68}, as
\begin{align}
	e(R, P_V) &= \sup_{\rho \geq 0} \Big\{ \rho R - E_s(\rho, P_V) \Big\},
\end{align}
where
\begin{align}
	E_s(\rho, P_V) &\triangleq \log \bigg( \sum_{v \in \mcl{V}} P_V(v)^{\frac{1}{1+\rho}}\bigg)^{1+\rho}.
\end{align}
The weak channel expurgated exponent is given by
\begin{align} \label{eq-weak-exp}
	\!E_{\mathrm{ex}}^{\prime}(Q, R) \!
	\triangleq \!\! \min_{\substack{P_{X\bar{X}}: P_X = Q \\ I_P(X; \bar{X}) \leq R}} \mathbb{E}_P\big[ d_B(X, \bar{X}) \big] \!+\! I_P(X; \bar{X})\! -\! R,
\end{align}
where $\!d_B(x, \bar{x}) \!\!\triangleq\!\! -\log \sum_y \sqrt{W(y|x) W(y|\bar{x})}$
is the Bhattacharyya distance.
This expression is referred to as ``weak'' since it is looser than the Csisz\'ar-K\"orner-Marton (CKM) expurgated exponent \cite{1056281}, given by
\begin{align}
	E_{\mathrm{ex}}(Q, R) 
	&\triangleq \!\!\! \min_{\substack{P_{X\bar{X}}\\ P_X=P_{\bar{X}}=Q\\I_P(X;\bar{X})\leq R}} \!\!\! \mathbb{E}_P\!\big[d_B(X,\bar{X}) \big] \!+\! I_P(X;\bar{X}) \! - \! R.
\end{align}
The dual expression of~\eqref{eq-weak-exp}, as given in~\cite[Eq.~9]{8006660}, is
\begin{align}
	E_{\mathrm{ex}}^{\prime}(Q, R) = \sup_{\rho \geq 1} E_{\mathrm{x}}^{\prime}({Q}, \rho) - \rho R,
\end{align}
where
\begin{align}
	E_{\mathrm{x}}^{\prime}({Q}, \rho) \triangleq
	\min _{Q^{\prime}}-\rho \sum_x Q(x) \log \sum_{\bar{x}} Q^{\prime}(\bar{x}) \, e^{-\frac{d_B(x, \bar{x})}{\rho}}.
\end{align}
%The Csisz\'ar JSCC expurgated exponent is given by
%\begin{align}
%	E_J^{\mathrm{ex}}(t, P_V) \triangleq \min_R \left[ t\,e\left(\frac{R}{t}, P_V \right) + \max_Q E_{\mathrm{ex}}^{\prime}(Q, R) \right].
%\end{align}
%Csisz\'ar~\cite{1056585} demonstrated the achievability of $E_J^{\mathrm{ex}}$ using the packing lemma~\cite[Lemma~6]{csiszar1980} and a partitioning of source sequences based on source type classes.
%Note that Csisz\'ar in \cite{1056585} obtained two achievable expurgated exponents. 
%However, in this paper, we focus only on the second expression, as it can only be achieved through partitioning, whereas the first one does not require such partitioning.
Using the packing lemma~\cite[Lemma~6]{csiszar1980}, Csisz\'ar~\cite{1056585} derived two achievable expurgated exponents for joint source-channel coding. 
% TODO: GIVE EXPRESSION. 
The first exponent was derived by assigning the same composition to all source messages, in a similar way to Gallager's~\cite[Problem~5.16]{gallager}  for the i.i.d. ensemble. 
The expression is given by
\begin{align} \label{eq-csiszar-org-1}
	E_{J, 1}^{\mathrm{ex}}(t, P_V) \triangleq  \max_Q \min_R \left[ te\!\left(\frac{R}{t}, P_V \right) +  E_{\mathrm{ex}}(Q, R) \right].
\end{align}
The second exponent requires partitioning of the source sequences and is given by
\begin{align} \label{eq-csiszar-org}
	E_{J, 2}^{\mathrm{ex}}(t, P_V) \triangleq \min_R \left[ te\!\left(\frac{R}{t}, P_V \right) + \max_Q E_{\mathrm{ex}}^{\prime}(Q, R) \right].
\end{align}
It is not known which of the two exponents is higher. 
%The question of which of the two exponents in~\cite{1056585} dominates the other remains open in general.
%Csisz\'ar partitions the source sequences according to their type (i.e. $m=N_k$), assigning a distinct codeword distribution to each one. 
%A more general form of partitioning was later introduced in~\cite{6803047}, where they derived the maximum a posteriori (MAP) random-coding error exponent for arbitrary partitioning of the source sequence set.
%Moreover, they showed that a particular partitioning of the source sequences into two classes can recover the Csisz\'ar random coding error exponent.
%The work in\cite{moeini2025} considers a less general partitioning, where each class contains one or more full source type classes, 
%and shows that this recovers the MAP random-coding error exponent of~\cite{6803047} using the universal maximum mutual information (MMI) decoder. In this work, we follow the partitioning approach of~\cite{moeini2025}. 
%We then show that two carefully chosen classes are also sufficient to achieve $E_J^{\mathrm{ex}}(t, P_V)$.
Csisz\'ar partitions the source sequences according to their type (i.e., $m = N_k$), assigning a distinct codeword distribution to each one. 
A more general partitioning was later proposed in~\cite{6803047}, where they derived the maximum a posteriori (MAP) random-coding error exponent for arbitrary partitioning of the source sequence set.
It was further shown that a specific two-class partitioning suffices to recover Csisz\'ar's random-coding exponent. 
%TODO-FINAL-MANUS: 
The work in~\cite{moeini2025dualdomainexponentmaximummutual} considers a less general partitioning, where each class contains one or more full source type classes, 
and shows that this recovers the MAP random-coding exponent of~\cite{6803047} using the universal maximum mutual information (MMI) decoder. 
In this work, we adopt the partitioning of~\cite{moeini2025dualdomainexponentmaximummutual}, and show that two carefully chosen classes are also sufficient to achieve $E_{J,2}^{\mathrm{ex}}(t, P_V)$.
%In this work, we consider a less general partitioning, where each class contains one or more full source type classes. 
%This setting includes Csisz\'ar's construction as a special case. We then show that two carefully chosen classes are sufficient to achieve $E_{J,2}^{\mathrm{ex}}(t, P_V)$.

The work in~\cite{8006660} revisits expurgation in joint source-channel coding 
%and rederives Csisz\'ar's exponents from~\cite{1056585} using Gallager's classical expurgation method~\cite[Sec.~5.7]{gallager}. 
and rederives Csisz\'ar's exponents from~\cite{1056585} using a new expurgation method.
The key result from\cite{8006660} is summarized in the following lemma.
\begin{lemma}\label{lem-expurgation-jscc}
	There exists a code $\mcl{C}$ such that for any source type $P_i \in \mcl{P}_k(\mcl{V})$ and any sequence $\bs{v} \in \mcl{T}^k(P_i)$, the error probability satisfies
	\begin{align} \label{eq-lem-exp}
		p_e(\bs{v}, {\mcl{C}})  \,\, \dot{\leq} \,\, 
		 \sum_{j=1}^{N_k} \, \mathbb{E}\Big[p_e\big(\bs{v}, j, \msf{C}\big)^{\frac{1}{\rho_{ij}}} \Big]^{\rho_{ij}},
	\end{align}
	where $\rho_{ij} > 0$, and $p_e\big(\bs{v}, j, \msf{C}\big)$ refers to the probability that, given $\bs{V} = \bs{v}$, there exists some $\bs{\bar{v}} \in \mcl{T}^k(P_j)$
	that yields a decoding metric at least as high as that of $\bs{v}$.\
	Moreover, the expectation is taken with respect to the given (not necessarily distinct) codeword distributions $\{P_{\bs{X}}^{(j)}\}_{j=1}^{N_k}$.
\end{lemma}
%We will use the code $\mcl{C}$ from Lemma~\ref{lem-expurgation-jscc} to derive the main results of this paper.
%TODO: I HAVE CHANGED NOTATIONS OF THE CODES HERE, THE PROOFS NEED TO BE UPDATED. (CODE WITH CALLIGRAPHIC FONT AND RANDOM CODE ENSEMBLE WITH SANSERIF FONT)

%%%%%%%%%%%%%%%%%%%%%%%%%%%%%%%%%%%%%%%%%%%%%%%%%%%%%%%%%%%%%%%%%%%%%%%%%%%%%%%%%%%%%%%%%%%%%
%%%%%%%%%%%%%%%%%%%%%%%%%%%%%%%%%%%%%%%%%%%%%%%%%%%%%%%%%%%%%%%%%%%%%%%%%%%%%%%%%%%%%%%%%%%%%
\section{Main Results} 
In this section, we present the main results. 
We begin with the results in the primal domain, followed by direct derivations of their counterparts in the dual domain;  their equivalence follows from Lagrange duality~\cite{Boyd_Vandenberghe_2004}. Direct dual-domain derivations have the advantages of remaining valid for arbitrary alphabets, and error exponents remain achievable for any choice of the optimization parameters.
% TODO: Primal vs Dual

In this paper, we consider the partitioning of the source sequences into $m$ disjoint sets $\mcl{A}_1, \ldots, \mcl{A}_m$, each containing one or more full type classes,
associated with a codeword distribution $Q_c \in \mathcal{P}_n(\mathcal{X})$, where the size of the type class must satisfy $\left|\mathcal{T}^n(Q_c)\right| \geq \left|\mcl{A}_c\right|$ for all $c = 1, \ldots, m$.
We denote by $\Lambda_c \subset \mcl{P}_k(\mcl{V})$ the set of source types included in $\mcl{A}_c$.
For each source sequence $\bs{v} \in \mcl{A}_c$, the codeword $\bs{x}_{\bs{v}} \in \mathcal{X}^n$ is drawn independently and uniformly from the type class $\mcl{T}^n(Q_c)$.
%where the size of the type class must satisfy $\left|\mathcal{T}^n(Q_c)\right| \geq \left|\mcl{A}_c\right|$ for all $c = 1, ..., m$.

\subsection{Primal-Domain}
% TODO: Mismatched Decoder
% TODO: Should this be lemma? or theorem
As a first step, we consider an arbitrary partition of the source sequences and derive an error bound for the code in Lemma~\ref{lem-expurgation-jscc}.
\begin{lemma} \label{lem-primal-given}
%	, where $|\mcl{T}^n(Q_c)| > |\mathcal{A}_c|$ for all $c = 1, \ldots, m$.
	Let $\mcl{A}_1,..., \mcl{A}_m$ be a given arbitrary partition of the source sequences, each associated with a codeword distribution $Q_c  \in  \mcl{P}_n(\mcl{X})$.
%	, each containing one or more full type classes. 
%	For each source sequence $\bs{v} \in \mcl{A}_c$, the codeword $\bs{x}(\bs{v}) \in \mathcal{X}^n$ is drawn independently and uniformly from the type class $\mcl{T}^n(Q_c)$, 
%	where the size of the type class must satisfy $\left|\mathcal{T}^n(Q_c)\right| \geq \left|\mcl{A}_c\right|$ for all $c = 1, \ldots, m$.
	Then there exists a codebook such that
	\begin{align}
		p_e \, \dot{\leq} 
		\sum_{c=1}^{m} \sum_{i \in \Lambda_c} \exp\!\left(\!-n\!\left[t{e}\!\left(\frac{R_i}{t}, P_V\!\right)\! +\! E_{\mathrm{ex}}^{\prime}(Q_c, R_i) \right]\!\right).
	\end{align}
\end{lemma}
The next theorem considers the partitioning that maximizes the exponent in Lemma~\ref{lem-primal-given}, for the given codeword distributions.

\begin{theorem} \label{thm-primal-best-partition}
	For a given set of codeword distributions $\mcl{Q}_m = \{ Q_1, Q_2, \ldots , Q_m \}$, 
	there exists a partition of the source sequences into $m$ disjoint sets such that the following exponent is achievable\
	\begin{align}
		\!\!\!{E}_{J, \mathrm{ex}}^{\mathrm{primal}}(t, \mcl{Q}_m)
		&\!\triangleq\min _{R}\!\left[t {e}\Big(\frac{{R}}{t},  P_V\Big) \!+\! \max _{{Q} \in \mcl{Q}_m} \!\! {E}_{\mathrm{ex}}^{\prime}({Q}, {R})\right] .
	\end{align}
\end{theorem}

\begin{remark}
	If  the distribution in $\mcl{Q}_m$ that maximizes $ {E}_{\mathrm{ex}}^{\prime}({Q}, {R})$ is independent of $R$, 
	then 
	\begin{align} \label{eq-remark-primal}
		\!\!{E}_{J, \mathrm{ex}}^{\mathrm{primal}}(t, \mcl{Q}_m)
		&\!=\!\!\max _{{Q}  \in \mcl{Q}_m} \! \min _{R}\!\left[t {e}\Big(\frac{{R}}{t},  P_V\Big)\! + \! {E}_{\mathrm{ex}}^{\prime}({Q}, {R})\right] .
	\end{align}
\end{remark}
Note that the exponent in \eqref{eq-remark-primal} can be achieved without partitioning, 
by generating all codewords from the type class of the distribution in $\mcl{Q}_m$ that maximizes ${E}_{\mathrm{ex}}^{\prime}({Q}, {R})$.
However, if it were known in advance that all codewords share the same composition,
%, as in the setting used by Csisz\'ar to derive $E_{J, 1}^{\mathrm{ex}}(t, P_V)$ given in \eqref{eq-csiszar-org-1},
then ${E}_{\mathrm{ex}}^{\prime}(Q, R)$ in~\eqref{eq-remark-primal} would be replaced by ${E}_{\mathrm{ex}}(Q, R)$, resulting in a higher exponent. 

%TODO: HOW DOES THIS RELATE TO THE FIRST EXPURGATED EXPONENT? 

\subsection{Dual-Domain}

% TODO: 
The following lemma presents the analogue of Lemma~\ref{lem-primal-given} in the dual domain.
%The following lemma presents the analogue of Lemma~\ref{lem-primal-given} in the dual domain, where the equivalent does not follow immediately from Lemma~\ref{lem-primal-given}.
\begin{lemma} \label{lem-dual-given}
	Consider a given partitioning of the source sequences as described in Lemma \ref{lem-primal-given}. 
	Then, there exists a codebook such that for any choice of parameters $\{\lambda_{i}, \rho_{i}\}$ satisfying $\lambda_{i} > 0$ and $\rho_{i} \geq 1$,
	\begin{align}\label{eq-exp-given-dual}
		\begin{split}
		&p_e \, \dot{\leq} \,
		\sum_{c=1}^{m}\sum_{i \in \Lambda_c} 
		\exp\!\bigg(\!\!-n\Big[{E}_{\mathrm{x}}^{\prime}\big({Q}_c, \rho_{i}\big) +\big(\lambda_{i}-\rho_{i}\big) {R}_i  \\
		&\hspace{3.5cm}-t {E}_{\mathrm{s}}\big(\lambda_{i}, P_V\big)\Big]\bigg).
		\end{split}
	\end{align}
\end{lemma}

%TODO: IS THIS RESULT ONLY VALID FO THE INF OVER LAMBDA/RHO? I SUPPOSE NOT, IN WHICH CASE, I WOULD SUGGEST TO STATE THE RESULT FOR ANY LAMBDA/RHO IN THE VALID RANGE.

The next corollary expresses the result of Lemma~\ref{lem-dual-given} in the same form as Theorem~1 in~\cite{6803047}.
% and Theorem~2 in~\cite{moeini2025}.

\begin{corollary}\label{cor-es}
	The expression in~\eqref{eq-exp-given-dual} can also be expressed as
	\begin{align}
		&\!p_e \, \dot{\leq} 
		\sum_{c=1}^{m} 
		\exp\!\left(\!-\! \sup_{\rho_c \geq 1}\! \left\{n{E}_{\mathrm{x}}^{\prime}\big({Q}_c, \rho_c\big) \!-\! tE_s^{(c)}\big(\rho_c, P^k\big) \right\}\!\right),
	\end{align}
	where
	\begin{align}
		 E_s^{(c)}\big(\rho, P^k\big) \triangleq \log \Big(\sum_{\bs{v}\in \mcl{A}_c} P^k\big(\bs{v}\big)^{\frac{1}{1+\rho}} \Big)^{1+\rho}.
	\end{align}
\end{corollary}

Next theorem considers the partitioning that maximizes the exponent in Lemma~\ref{lem-dual-given}.

\begin{theorem} \label{thm-dual-best-partition}
	Consider a given set of codeword distributions $\mcl{Q}_m = \{ Q_1, Q_2, \ldots, Q_m \}$, as described in Theorem~\ref{thm-primal-best-partition}. 
	Then, there exists a partition of the source sequences into $m$ disjoint sets such that the following exponent is achievable\
	\begin{align} \label{eq-exp-dual}
		&{E}_{J, \mathrm{ex}}^{\mathrm{dual}}(t, \mcl{Q}_m)
		\triangleq \sup _{\lambda\geq 1}\Big\{\overline{E}^{\prime}_{\mathrm{x}}\big(\mcl{Q}_m, \lambda\big)-t{E}_{\mathrm{s}}\big(\lambda, P_V\big)\Big\},
	\end{align}
	where 
	\begin{align}
		\overline{E}^{\prime}_{\mathrm{x}}\big(\mcl{Q}_m, \lambda\big) \triangleq \min_R \sup_{\rho \geq 1} \, E_{\mathrm{x}}^{\prime}(\mcl{Q}_m, \rho) + \big(\lambda  - \rho\big)R,
	\end{align}
	and
	$E_{\mathrm{x}}^{\prime} (\mcl{Q}_m, \rho) \triangleq \max_{Q \in \mcl{Q}_m} {E}_{\mathrm{x}}^{\prime} (Q, \rho)$.
\end{theorem}

% TODO: IT WOULD BE GOOD TO HAVE $\overline{E}_x'$ RATHER THAN $\overline{E_x'}$

%The next theorem 
%It can be seen that if the zero-error capacity of the channel $W$ is zero, then $\overline{{E}_{\mathrm{x}}^{\prime}}\big(\lambda,\mcl{Q}_m\big)$ is finite.
%TODO-FINAL-MANUS: 
%The equivalence of Theorem~\ref{thm-primal-best-partition} and Theorem~\ref{thm-dual-best-partition} can be seen using Lagrange duality.
%The next result extends~\cite[Lemma 3]{1614076} to the expurgated exponent for a given set of codeword distributions.
%The equivalence between Theorems~\ref{thm-primal-best-partition} and~\ref{thm-dual-best-partition} follows from Lagrange duality. 
%TODO-FINAL-MANUS:
Theorem~\ref{thm-dual-best-partition} is the dual counterpart of Theorem~\ref{thm-primal-best-partition}, and their equivalence follows from Lagrange duality.
The following result extends~\cite[Lemma 3]{1614076} to the expurgated exponent for a given set of codeword distributions.

\begin{theorem}\label{theorem-dual-concave-hull}
	The function $\overline{E}_{\mathrm{x}}^{\prime}(\mcl{Q}_m, \rho)$
	is the concave hull on the interval $[1, \infty)$ of the function $E_{\mathrm{x}}^{\prime}(\mcl{Q}_m, \rho)$.
%	TODO: DEFINED IN (XX).
\end{theorem}

\begin{remark}
	If $E_{\mathrm{x}}^{\prime}(\mcl{Q}_m, \rho)$ is concave in $\rho$, 
	or equivalently, if the distribution in $\mcl{Q}_m$ that maximizes $E_{\mathrm{x}}^{\prime}(\rho, Q)$ is independent of $\rho$, 
	then $\overline{E}_{\mathrm{x}}^{\prime}(\mcl{Q}_m, \rho) = E_{\mathrm{x}}^{\prime}(\mcl{Q}_m, \rho)$, and hence
	\begin{align} \label{eq-remark-dual}
		E_{J, \mathrm{ex}}^{\mathrm{dual}}(t, \mcl{Q}_m)
		= \sup_{\rho \geq 1} \Big\{ E_{\mathrm{x}}^{\prime}(\mcl{Q}_m, \rho) - tE_{\mathrm{s}}(\rho, P_V) \Big\}.
	\end{align}
\end{remark}
%TODO: The equivalence of~\eqref{eq-remark-primal} and~\eqref{eq-remark-dual} follows from Lagrange duality~\cite{Boyd_Vandenberghe_2004}. 
%TODO-FINAL-MANUS:
The next corollary considers Csisz\'ar's setting, where the partitioning is based on source types, and presents the corresponding dual expression of $E_{J, 2}^{\mathrm{ex}}(t, P_V)$ given in~\eqref{eq-csiszar-org}.
\begin{corollary} \label{cor-csiszar-dual}
%	TODO: WHAT EXACTLY DOES THE CONDITION MEAN? 
%	If the partitioning is based on source type classes and the associated distributions are chosen appropriately, then
	If the partitioning is based on source type and the associated distributions $\{Q_i\}_{i=1}^{N_k}$ are chosen according to
	\begin{align}
%		\begin{split}
%			&Q_i = \underset{\tilde{Q}}{\operatorname{argmax}}\, \sup_{\rho \geq 1}\,  \sup_{\lambda > 0} \, {E}_{\mathrm{x}}^{\prime}\big(\tilde{Q}, \rho\big) \\
%			&\hspace{2.75cm}+ \big(\lambda - \rho\big)R_i -t{E}_{\mathrm{s}}\big(\rho, P_V\big),
%		\end{split}
	\!\!\!Q_i\! = \underset{\tilde{Q}}{\operatorname{argmax}}\!\!\! \sup_{\rho \geq 1, \lambda > 0}\!\! {E}_{\mathrm{x}}^{\prime}\big(\tilde{Q}, \rho\big)\!+\!\big(\lambda \!-\! \rho\big)R_i \!-\!t{E}_{\mathrm{s}}\big(\lambda, P_V\big),
	\end{align}
	then
	\begin{align} \label{eq-csiszar-dual}
		E_{J, \mathrm{ex}}^{\mathrm{dual}}\big(t, \{Q_i\}\big) = \sup_{\lambda \geq 1} \Big\{\overline{{E}}_{\mathrm{x}}^{\prime}\big(\lambda\big)-t{E}_{\mathrm{s}}\big(\lambda, P_V\big)\Big\},
	\end{align}
	where $\overline{{E}}_{\mathrm{x}}^{\prime}\big(\lambda\big)$ is the concave hull of 
	${E}_{\mathrm{x}}^{\prime}\big(\lambda\big) \triangleq \max_ Q{E}_{\mathrm{x}}^{\prime}\big(Q,\! \lambda\big)$. 
%	TODO: THE LAST OPTIMIZATION OVER Q IS FOR $Q\in \mathcal{Q}_m$?
\end{corollary}
%TODO: It can be seen that~\eqref{eq-csiszar-dual} is the dual expression of Csisz\'ar's exponent $E_{J,2}^{\mathrm{ex}}(t, P_V)$ given in~\eqref{eq-csiszar-org}.

%The result in Corollary~\ref{cor-csiszar-dual} appeared in~\cite{8006660} as the dual expression of Csisz\'ar's expurgated exponent $E_J^{\mathrm{ex}}(t, P_V)$. 
%However, the expression was not derived directly; instead, it was obtained using Lagrange duality.

% TODO: the work by jonathan had stated this, but for source type partitioning and just using duality and not direct! The approach can actually be used to obtain theorem of random coding directly!
%A weaker version of Theorem~\ref{theorem-dual-concave-hull} was presented in~\cite{8006660}, 
%where the partitioning is based on source type classes and the result was derived indirectly using Lagrange duality. 
%If the partitioning is based on source type classes, it was shown in
%Here, by contrast, we obtain the result directly in the dual domain.

\subsection{Two-Class Partition of the Source Sequences}
Now, we assume that we only have two classes, i.e. $m=2$.
%TODO-FINAL-MANUS:
\begin{lemma}\label{lem-partition}
	For a pair of distributions $\mcl{Q} = \{ Q, Q^{\prime} \}$,
	the partition that achieves $E_{J, \mathrm{ex}}^{\mathrm{dual}}(t,\mcl{Q})$ (or $E_{J, \mathrm{ex}}^{\mathrm{primal}}(t,\mcl{Q})$), 
	is 
	\begin{align}
		\mcl{A}_1 &= \left\{ \bs{v} : \big| \mcl{T}^k(\hat{P}_{\bs{v}}) \big| \leq e^{nR_0} \right\} \\
		\mcl{A}_2 &= \left\{ \bs{v} : \big| \mcl{T}^k(\hat{P}_{\bs{v}}) \big| > e^{nR_0} \right\},
	\end{align}
	for some $R_0 > 0$ that depends on $t, P_V$, $W$, and $\mcl{Q}$.
\end{lemma}

%The following corollary is the expurgated analogue of~\cite[Corollary~1]{6803047}. 
%As in the random-coding case, a direct application of Carathéodory's theorem \cite[Cor. 17.1.5]{rockafellar1970a} implies that any point on the graph of $\overline{E}_{\mathrm{x}}^{\prime}(\lambda)$ can be expressed as a convex combination of two points on the graph of ${E_{\mathrm{x}}^{\prime}}(\lambda)$. 
% Assume the first point is associated with $Q$ and the second point
%As a result, there exists a pair of distributions $\{Q, Q'\}$ such that both points lie on the graph of ${E_{\mathrm{x}}^{\prime}}(\lambda, \{Q, Q'\})$. Optimizing over all such pairs gives the following result.
%TODO-FINAL-MANUS: 
The following corollary is the expurgated analogue of~\cite[Corollary~1]{6803047}. 
As in the random-coding case, a direct application of Carathéodory's theorem \cite[Cor. 17.1.5]{rockafellar1970a} 
implies that for any $\lambda_0$, the value $\overline{E}_{\mathrm{x}}^{\prime}(\lambda_0)$ can be exprssed as a convex combination of two points on the graph of $E_{\mathrm{x}}^{\prime}(\lambda)$. 
Let these points correspond to $\lambda_1$ and $\lambda_2$, and let $Q_1$ and $Q_2$ be the respective maximisers of $E_{\mathrm{x}}^{\prime}(\tilde{Q}, \lambda)$ at $\lambda_1$ and $\lambda_2$.
It follows that both points also lie on the graph of $E_{\mathrm{x}}^{\prime}(\{Q_1, Q_2\},\lambda)$, and thus $\overline{E}_{\mathrm{x}}^{\prime}(\lambda_0)$ also lies on its concave hull.
This shows that a two-class partitioning suffices to achieve $\overline{E}_{\mathrm{x}}^{\prime}(\lambda_0)$.
Optimizing over all such pairs gives the following result.
%implies that for any $\lambda_0$, $\overline{E}_{\mathrm{x}}^{\prime}(\lambda_0)$ can be expressed as a convex combination of two points on the graph of ${E_{\mathrm{x}}^{\prime}}(\lambda)$. Assume the first is associated with $\lambda_1$ and the second with $\lambda_2$, where $Q_1$ and $Q_2$ are respectively the solutions to
%$\max_{\tilde{Q}}{E_{\mathrm{x}}^{\prime}}(\tilde{Q}, \lambda)$ for $\lambda_1$ and $\lambda_2$ respectively.

%TODO-FINAL-MANUS: 
\begin{corollary}\label{cor-main}
	There exists a partition of the source sequences into two classes, 
%	each associated with a distribution from a given pair, 
	such that $E_{J,2}^{\mathrm{ex}}(t, P_V)$ in~\eqref{eq-csiszar-org} is achievable.
\end{corollary}
An immediate implication of Corollary~\ref{cor-main} is that $E_{J,2}^{\mathrm{ex}}(t, P_V)$ can be achieved with only two classes, in contrast to Csisz\'ar's construction~\cite{1056585}, 
which requires $m = N_k$ classes, with $N_k$ growing polynomially in $k$.

%TODO-FINAL-MANUS: 
%\section{In the Absence of Partitioning}
\section{Single Class Coding}

%In this section, we discuss only the dual-domain counterpart of our results in the absence of partitioning, where all codewords are assumed to have the same composition.
%This section presents the results in the absence of partitioning, where all codewords share the same composition. We focus on the dual-domain counterpart of the results.
This section presents the dual-domain counterparts of our results in the absence of partitioning, where all codewords share the same composition.
%The case of no partitioning in JSCC can be attributed to Gallager, as the random-coding exponent he derived in~\cite[Problem~5.16]{gallager} does not consider any partitioning of the source sequences.
%We attribute the non-partitioned setting in JSCC to Gallager, as the random-coding exponent he derives in~\cite[Problem~5.16]{gallager} does not involve any partitioning of the source sequences.
%The following lemma presents the analogue of Lemma~\ref{lem-primal-given} in the absence of partitioning. 
%The following lemmas presents the analogues of Lemma~\ref{lem-dual-given} and Theorem~\ref{thm-dual-best-partition} in this setting.
%The following Lemmas present the analogues of Theorem~\ref{thm-dual-best-partition} and Corollary~\ref{cor-csiszar-dual} in this setting.
%
The following lemmas present the counterparts of Theorem~\ref{thm-dual-best-partition} and Corollary~\ref{cor-csiszar-dual} in this setting
\begin{lemma} \label{lem-primal-nopart-opt}
	There exist single-class codes with constant composition ensembles that satisfies
	\begin{align} 
		p_e \,\,\dot{\leq} \, \exp\left(-n \left[\, \sup _{\rho\geq 1}\Big\{{{E}_{\mathrm{x}}}\big(Q,\rho\big)-t{E}_{\mathrm{s}}\big(\rho, P_V\big)\Big\} \right] \right),
	\end{align}
	where 
	\begin{align}
		\begin{split}
			&{E}_{\mathrm{x}}(Q,\rho)\\
			&\triangleq \sup_{a(.)} -\rho \sum_x Q(x) \log \sum_{\bar{x}} Q(\bar{x})\! \left(e^{-d_B(x, \bar{x})}\frac{e^{a(\bar{x})}}{e^{{a(x)}}} \right)^{\frac{1}{\rho}}.
		\end{split}
	\end{align} 
\end{lemma}

\begin{corollary} \label{cor-primal-nopart-opt}
 	If the codeword distribution in Lemma~\ref{lem-primal-nopart-opt} is chosen as
	\begin{align}
		Q = \underset{\tilde{Q}}{\operatorname{argmax}} \, \bigg\{ \sup_{\rho \geq 1} \, {E}_{\mathrm{x}}\big(\tilde{Q}, \rho\big) - t{E}_{\mathrm{s}}\big(\rho, P_V\big)\bigg\},
	\end{align}
	then there exists a codebook achieving the following exponent
	\begin{align} \label{eq-no-part-best-dual}
		E_{G, \mathrm{ex}}^{\mathrm{dual}}(t) 
		&\triangleq \sup _{\rho\geq 1}\Big\{{{E}_{\mathrm{x}}}(\rho)-t{E}_{\mathrm{s}}\big(\rho, P_V\big)\Big\} ,
	\end{align}
	where ${E}_{\mathrm{x}}(\rho) \triangleq \max_Q {E}_{\mathrm{x}}(Q, \rho)$.
\end{corollary}
The expression in~\eqref{eq-no-part-best-dual} is the dual counterpart of $E_{J, 1}^{\mathrm{ex}}(t, P_V)$ given in~\eqref{eq-csiszar-org-1}.
This result was previously derived by Gallager in~\cite[Problem~5.16]{gallager}, where the i.i.d. ensemble was used instead of constant-composition coding.

The result of Corollary~\ref{cor-primal-nopart-opt} cannot be directly compared with Corollary~\ref{cor-csiszar-dual}, 
since the concave hull of $E_{\mathrm{x}}^{\prime}(\lambda)$ can, in general, be larger than $E_{\mathrm{x}}(\lambda)$, 
even though $E_{\mathrm{x}}(\lambda) \geq E_{\mathrm{x}}^{\prime}(\lambda)$ always holds.
Thus, whether partitioning can generally improve the achievable expurgated exponent remains an open question, as pointed out in~\cite{1056585, 8006660}.
This contrasts with the case of the random-coding exponent, where it is known that partitioning can strictly improve the exponent \cite{1614076, 6803047}.

\bibliographystyle{IEEEtran}
\bibliography{refs}

\newpage

\section{Proofs}
% TODO: instead of proof, I use <derivation>
Due to space constraints, we present only the direct derivations of the dual-domain results, as they are less straightforward and not as immediate as those in the primal domain. 
The primal-domain case follows by similar arguments, using standard tools from the method of types.

\subsection{Proof of Lemma \ref{lem-dual-given}}
The proof closely follows the approach in~\cite[Sec.~III.C]{8006660}.
We consider the code from Lemma~\ref{lem-expurgation-jscc}, and use the following decoding metric
\begin{align}
	q(\bs{v}, \bs{x}, \bs{y})=W^n(\bs{y} | \bs{x}) \exp\!\big(-2 k D(P_i \| P_V)\big)
\end{align}
if $\bs{v} \in \mcl{T}^k(P_i)$.
Then, for any $s > 0$, we can upper bound $p_e\big(\bs{v}, j, \msf{C}\big)$ as
\begin{align}
	\!\!p_e\big(\bs{v}, j, \msf{C}\big)\!
	&= \mathbb{P}\!\left[\bigcup_{\bar{\bs{v}}\in \mcl{T}^k(P_j)} \!\!\! q(\bs{\bar{v}}, \bs{x}_{\bar{\bs{v}}}, \bs{Y})  \geq  q(\bs{v}, \bs{x}_{\bs{v}}, \bs{Y}) \right]\\
	&= \mathbb{P}\!\left[\bigcup_{\bar{\bs{v}}\in \mcl{T}^k(P_j)} \!\!\! q(\bs{\bar{v}}, \bs{x}_{\bar{\bs{v}}}, \bs{Y})^s  \geq  q(\bs{v}, \bs{x}_{\bs{v}}, \bs{Y})^s \right]\\
	& \leq \sum_{\bs{y}} W^n(\bs{y}|\bs{x}_{\bs{v}}) \!\! \sum_{\bar{\bs{v}}\in \mcl{T}^k(P_j)} \left(\frac{q(\bs{\bar{v}}, \bs{x}_{\bar{\bs{v}}}, \bs{y})}{q(\bs{v}, \bs{x}_{\bs{v}}, \bs{y})} \right)^s.
\end{align}
Substituting the decoding metric and setting $s = \frac{1}{2}$, we obtain
\begin{align}
	\begin{split}
		p_e\big(\bs{v}, j, \msf{C}\big)\,
		&\leq \left(\frac{\exp \left(-k D\left(P_j \| P_V\right)\right)}{\exp \left(-k D\left(P_i \| P_V\right)\right)}\right) \\
		&\times \sum_{\bar{\bs{v}} \in \mcl{T}^k(P_j)} \sum_{\bs{y}} \sqrt{W^n(\bs{y} | \bs{x}_{\bs{v}}) W^n(\bs{y} | \bs{x}_{\bar{\bs{v}}})}.
	\end{split}
\end{align}
If we assume $\rho_{ij} \geq 1$ for all $(i, j)$, we can use Hölder's inequality \cite[Problem~4.15.f]{gallager} to move the summation over $\bs{\bar{v}}$ outside, as follows
\begin{align}
	\begin{split}
		\!\!p_e\big(\bs{v}, j, \msf{C}\big)^{\frac{1}{\rho_{ij}}} \!&\leq 
		\left(\frac{\exp (-k D(P_j \| P_V))}{\exp(-k D(P_i \| P_V))}\right)^{\frac{1}{\rho_{i j}}}\\
		&\times \!\!\!\!\! \sum_{\bar{\bs{v}} \in \mathcal{T}(P_j)}  \prod_{l=1}^n\!\left[\sum_y\! \sqrt{W(y | x_{\bs{v}}^l) W(y | x_{\bs{\bar{v}}}^l)}\right]^{\frac{1}{\rho_{i j}}}\!\!\!,
	\end{split}
\end{align}
where $x_{\bs{v}}^l$ denotes the $l$-th element of $\bs{x}_{\bs{v}}$.
We now take the expectation of the above expression
\begin{align}
	\begin{split}
		\mathbb{E}\left[p_e\big(\bs{v}, j, \msf{C}\big)^{\frac{1}{\rho_{ij}}}\right] \,&\leq 
		\left(\frac{\exp (-k D(P_j \| P_V))}{\exp(-k D(P_i \| P_V))}\right)^{\frac{1}{\rho_{i j}}}\\
		&\times \mathbb{E}\left[\sum_{\bar{\bs{v}} \in \mathcal{T}^k(P_j)}  \prod_{l=1}^n \, e^{\frac{-d_B(X_l,  \bar{X}_l)}{\rho_{ij}}}\right],
	\end{split}
\end{align}
where the expectation is taken with respect to $P_{\bs{X}}^{(\hat{i})} \times P_{\bs{X}}^{(\hat{j})}$, where $P_{\bs{X}}^{(c)}$ denotes the uniform distribution over the type class $\mcl{T}^n(Q_c)$.
 Here, $\hat{i}$ denotes the index of the class to which source type $i$ belongs.
We can evaluate this expectation as follows
\begin{align}
	\begin{split}
		&\mathbb{E}\left[\, \prod_{l=1}^n \, e^{\frac{-d_B(X_l,  \bar{X}_l)}{\rho_{ij}}}\right]\\
		&\hspace{1cm}= \sum_{(\bs{x}, \bar{\bs{x}})} P_{\bs{X}}^{(\hat{j})}({\bs{x}}) P_{\bs{X}}^{(\hat{i})}(\bar{\bs{x}})\prod_{k=1}^n \, e^{\frac{-d_B(x_l,  \bar{x}_l)}{\rho_{i j}}}.
	\end{split}
\end{align}
We use  \cite[Lemma~2.1]{poltyrev} to rewrite the first expectation as if the distribution were i.i.d. with $Q_{\hat{i}}$. 
This gives
\begin{align}
	\begin{split}
		&\mathbb{E}\left[\, \prod_{l=1}^n \, e^{\frac{-d_B(X_l,  \bar{X}_l)}{\rho_{ij}}}\right]\\
		&\hspace{0.5cm}\,\dot{\leq}\sum_{\bs{x} \in \mcl{T}^n(Q_{\hat{j}} )} \frac{1}{\big|\mcl{T}^n(Q_{\hat{j}}) \big|} \prod_{l=1}^n \sum_{\bar{x}} Q_{\hat{i}}(x) \, e^{\frac{-d_B(x_l,  \bar{x})}{\rho_{i j}}}
%		\,\, \sum_{\bs{x}} P_{\bs{X}}^{(\hat{j})}({\bs{x}}) \prod_{k=1}^n \sum_{\bar{x}} Q_{\hat{i}}(x) \, e^{\frac{-d_B(x_k,  \bar{x})}{\rho_{i j}}}
	\end{split}\\
%	&\hspace{0.5cm}= \sum_{\bs{x} \in \mcl{T}^n(Q_{\hat{j}} )} \frac{1}{\big|\mcl{T}^n(Q_{\hat{j}}) \big|} \prod_{k=1}^n \sum_{\bar{x}} Q_{\hat{i}}(x) \, e^{\frac{-d_B(x_k,  \bar{x})}{\rho_{i j}}}\\
	&\hspace{0.5cm}\doteq \, \prod_{x \in \mcl{X}} \left[ \sum_{\bar{x}} Q_{\hat{i}}(x) \, e^{\frac{-d_B(x,  \bar{x})}{\rho_{i j}}}\right]^{nQ_{\hat{j}}(x)}\\
	&\hspace{0.5cm} = \exp\left(n \sum_x Q_{\hat{j}}(x) \log \sum_{\bar{x}} Q_{\hat{i}}(x) \, e^{\frac{-d_B(x,  \bar{x})}{\rho_{i j}}}  \right). \label{eq-does-not-depend}
\end{align}
Note that the bound in~\eqref{eq-does-not-depend} does not depend on $\bs{\bar{v}}$, and so
\begin{align}
	\begin{split}
		&\,\mathbb{E}\left[\sum_{\bar{\bs{v}} \in \mathcal{T}^k(P_j)}  \prod_{k=1}^n \, e^{\frac{-d_B(X_k,  \bar{X}_k)}{\rho_{ij}}}\right]\\
		&\,\,\dot{\leq} \exp\!\left(-n\! \left[-\sum_x Q_{\hat{j}}(x) \log \sum_{\bar{x}} Q_{\hat{i}}(x) \, e^{\frac{-d_B(x,  \bar{x})}{\rho_{i j}}}\!\!-R_j \right] \right).
	\end{split}
\end{align}
% TODO: changed
Raising both sides to the power $\rho_{ij}$ and taking the worst-case choice of $Q_{\hat{i}}$, we obtain
\begin{align}
	&\mathbb{E}\left[p_e\big(\bs{v}, j, \msf{C}\big)^{\frac{1}{\rho_{ij}}}\right]^{\rho_{ij}}\nonumber \\
%	\begin{split}
%		&\mathbb{E}\left[p_e\big(\bs{v}, j, \mcl{C}\big)^{\frac{1}{\rho_{ij}}}\right]^{\rho_{ij}}\\
%		&\dot{\leq}\,\,\, \left(\frac{\exp (k D(P_j \| P_V))}{\exp(k D(P_i \| P_V))}\right)\\
%		&\times \exp\!\bigg(\!\!\!-\!n\!\bigg[\!\!-\!\rho_{ij}\!\sum_x \! Q_{\hat{j}}(x) \log \sum_{\bar{x}}\! Q_{\hat{i}}(x) e^{\frac{-d_B(x,  \bar{x})}{\rho_{i j}}}\!\!\!\!\!-\rho_{ij}R_j \bigg] \bigg)
%	\end{split}\\
	&\leq \! \left(\frac{\exp (-k D(P_j \| P_V))}{\exp(-k D(P_i \| P_V))}\right)  \exp\!\left(\!-n\! \left[E_{\mathrm{x}}^{\prime}({Q_{\hat{j}}}, \rho_{ij})\!-\!\rho_{ij}R_j \right]\right).
\end{align}
Having evaluated the expectation in~\eqref{eq-lem-exp}, we now obtain the following bound on the error probability of the code in Lemma~\ref{lem-expurgation-jscc} as follows
\begin{align}
	p_e(\mcl{C}) 
	&= \sum_{\bs{v}} P^k(\bs{v})\, p_e(\bs{v},\mcl{C})\\
	\begin{split}
		&\,\, \dot{\leq}\,\, \sum_{(i,j)} \sum_{\bs{v}\in \mcl{T}^k(P_i)} P^k(\bs{v})
		\left(\frac{\exp (-k D(P_j \| P_V))}{\exp(-k D(P_i \| P_V))}\right) \\
		&\hspace{1cm} \times \exp\left(-n \left[E_{\mathrm{x}}^{\prime}({Q_{\hat{j}}}, \rho_{ij}) - \rho_{ij}R_j \right]\right). \label{eq-bound-rhoij}
	\end{split}
\end{align}
Observe that $ \sum_{\bs{v}\in \mcl{T}^k(P_i)} P^k(\bs{v}) \doteq \exp(-k D(P_i \| P_V))$. This implies that
\begin{align} \label{eq-to-be-ref-1}
	\!\!\!\!\!\!\!\!\sum_{\bs{v}\in \mcl{T}^k(P_i)} \!\!\!\!\! P^k(\bs{v})\! \left(\frac{\exp (-k D(P_j \| P_V))}{\exp(-k D(P_i \| P_V))}\right)
	&\!\doteq \exp (-k D(P_j \| P_V)).
\end{align}
We can further simplify this for any $\lambda_j > 0$ as follows
\begin{align}
	\begin{split}
		&\exp \big(-k D(P_j \| P_V)\big) \\
		&\hspace{0.25cm}=\exp \left(k \sum_v P_j(v) \log \frac{P_V(v)}{P_j(v)}\right)
	\end{split}\\
%	&\hspace{0.25cm}=\exp \left(k (1+\lambda_j) \sum_v P_j(v) \log \frac{P_V(v)^{\frac{1}{1+\lambda_j}}}{P_j(v)^{\frac{1}{1+\lambda_j}}}\right)\\
	&\hspace{0.25cm}=\exp\! \left(k (1+\lambda_j) \sum_v P_j(v) \log \frac{P_V(v)^{\frac{1}{1+\lambda_j}}}{P_j(v) P_j(v)^{\frac{-\lambda_j}{1+\lambda_j}}}\right).
\end{align}
The last expression can be written as
\begin{align}
	\begin{split}
		&\exp \left(k (1\!+\!\lambda_j) \sum_v P_j(v) \log \frac{P_V(v)^{\frac{1}{1+\lambda_j}}}{P_j(v) P_j(v)^{\frac{-\lambda_j}{1+\lambda_j}}}\right)\\
		&=\exp\!\left(\!n\!\left[t (1\!+\!\lambda_j)\! \sum_v P_j(v) \log \frac{P_V(v)^{\frac{1}{1+\lambda_j}}}{P_j(v)} \!-\! \lambda_jR_j\right]\right)
	\end{split}\\
	&\leq \exp\!\left(\!n\!\left[t (1\!+\!\lambda_j)\! \log \sum_v P_j(v) \frac{P_V(v)^{\frac{1}{1+\lambda_j}}}{P_j(v)} \!-\! \lambda_jR_j\right]\right) \label{eq-applied-jensen} \\
	&= \exp\left(-n\Big[\lambda_j R_j \!-\! tE_s(\lambda_j, P_V) \Big] \right),
\end{align}
where we applied Jensen's inequality in~\eqref{eq-applied-jensen}. 
By setting $\rho_{ij} = \rho_j$, we can upper bound the expression in~\eqref{eq-bound-rhoij} as
\begin{align}
	\begin{split}
		&p_e(\mcl{C}) \,\,\dot{\leq}\,\, \sum_{j=1}^{N_k} \exp\bigg(-n\Big[ {E}_{\mathrm{x}}^{\prime}\big({Q}_{\hat{j}}, \rho_j\big)\\
		&\hspace{2cm}+\big(\lambda_j-\rho_j\big) {R}_j-t \mathrm{E}_{\mathrm{s}}\big(\lambda_j, P_V\big) \Big]\bigg)
	\end{split}\\
	\begin{split}
		&\hspace{1cm}=\sum_{c=1}^{m} \sum_{i \in \Lambda_c} \exp\bigg(-n\Big[ {E}_{\mathrm{x}}^{\prime}\big({Q}_{c}, \rho_{i}\big)\\
		&\hspace{2cm}+\big(\lambda_{i}-\rho_{i}\big) {R}_{i}-t \mathrm{E}_{\mathrm{s}}\big(\lambda_{i}, P_V\big) \Big]\bigg).
	\end{split}
\end{align}
%Minimizing over $\rho_{ic}$ and $\lambda_{ic}$ concludes the proof.
To prove Corollary~\ref{cor-es}, observe that by applying~\eqref{eq-to-be-ref-1}, the bound in~\eqref{eq-bound-rhoij} can be rewritten as
\begin{align}
	p_e(\mcl{C})
	&\, \dot{\leq}\, \sum_{i=1}^{N_k} \sum_{\bs{v} \in \mcl{T}^k(P_i)} \frac{P^k(\bs{v})}{\hat{P}_{\bs{v}}(\bs{v})^{\rho_i}} \exp\big(-n\,{E}_{\mathrm{x}}^{\prime}(Q_{\hat{i}}, \rho_{i})\big) \\
	&= \sum_{c=1}^{m} \left( \sum_{\bs{v} \in \mcl{A}_c} \frac{P^k(\bs{v})}{\hat{P}_{\bs{v}}(\bs{v})^{\rho_{c}}} \right) \exp\!\big(\!-\!n\,{E}_{\mathrm{x}}^{\prime}(Q_c, \rho_{c})\big).
\end{align}
The inner sum can be upper bounded as
\begin{align}
	\!\!\!\! \sum_{\bs{v} \in \mcl{A}_c}\frac{P^k(\bs{v})}{\hat{P}_{\bs{v}}(\bs{v})^{\rho}} 
	&= \sum_{i \in \Lambda_c} \sum_{\bs{v} \in \mcl{T}^k(P_i)} \frac{P^k(\bs{v})}{P_i^k(\bs{v})^{\rho}} \\
	&\doteq \sum_{i \in \Lambda_c} \exp\left(k\sum_v P_i(v) \log \frac{P_V(v)}{P_i(v)^{1+\rho}}\right) \label{eq-card-source-class} \\
	&= \sum_{i \in \Lambda_c} \!\left[\exp\!\left(\!k\!\sum_v \! P_i(v)\! \log\! \frac{P_V(v)^{\frac{1}{1+\rho}}}{P_i(v)}\!\right)\!\right]^{1+\rho}\\
	&= \sum_{i \in \Lambda_c} \left[\sum_{\bs{v} \in \mcl{A}_c} P^k(\bs{v})^{\frac{1}{1+\rho}}\right]^{1+\rho}\\
	&\leq \left[\sum_{i \in \Lambda_c}  \sum_{\bs{v} \in \mcl{A}_c} P^k(\bs{v})^{\frac{1}{1+\rho}}\right]^{1+\rho} \label{eq-gal-holder-f} \\
	&= \left[\sum_{\bs{v} \in \mcl{A}_c} P^k(\bs{v})^{\frac{1}{1+\rho}}\right]^{1+\rho},
\end{align}
where in \eqref{eq-card-source-class}, we used the fact that $|\mcl{T}^k(P_i)| \doteq \exp(kH(P_i))$, and in \eqref{eq-gal-holder-f}, 
we applied the inequality $\sum_i a_i^r \leq (\sum_i a_i)^r$ for $r \geq 1$ \cite[Problem~4.15.f]{gallager}.
%\begin{align}
%        \sum_{\bs{v} \in \mcl{A}_c} \frac{P^k(\bs{v})}{\hat{P}_{\bs{v}}(\bs{v})^{\rho}} 
%        &= \sum_{i \in \Lambda_c} \Bigg[\sum_{\bs{v} \in \mcl{T}^k(P_i)} \!\!\!\! P^k(\bs{v})^{\frac{1}{1+\rho}} \Bigg]^{1+\rho}\\
%        &\leq \Bigg[\sum_{\bar{P}\in \Lambda_i}  \sum_{\bs{v} \in \mcl{T}^k(\bar{P})} \!\!\!\! P^k(\bs{v})^{\frac{1}{1+\rho_i}} \Bigg]^{1+\rho_i} 
%        &= \Bigg[ \sum_{\bs{v} \in \mcl{A}_k^{(i)}} P^k(\bs{v})^{\frac{1}{1+\rho_i}} \Bigg]^{1+\rho_i},
%    \end{align}
%\begin{align}
%	\sum_{\bs{v} \in \mcl{A}_c} \frac{P^k(\bs{v})}{\hat{P}_{\bs{v}}(\bs{v})^{\rho_{c}}} \leq E_s^{(c)}\big(\rho_{c}, P^k\big),
%\end{align}
%This completes the proof.

%%%%%%%%%%%%%%%%%%%%%%%%%%%%%%%%%%%%%%%%%%%%%%%%%%%%%%%%%%%%%%%%%%%%%%%%%%%%%%%%%%%%%%%%%%%%%%%%%%%%

\subsection{Proof of Theorem \ref{thm-dual-best-partition}}
To find the optimal partitioning, assume each class is associated with given values of $\rho_i$ and $\lambda_i$, which will be optimized later.
The source type class $\mcl{T}^k(P_i)$ is then assigned to class $c^\star$ if
\begin{align}\label{eq-best-partition}
	c^{\star}=\underset{c \in\{1, \ldots, m\}}{\operatorname{argmax}} E_{\mathrm{x}}^{\prime}(\mathrm{Q}_c, \rho_i).
\end{align}
This means that
\begin{align}
	\begin{split}
		&p_e(\mcl{C}) \,\, \dot{\leq}\,\, \sum_{j=1}^{N_k} \exp\bigg(-n\bigg[\max_{Q \in \mcl{Q}_m} \, {E}_{\mathrm{x}}^{\prime}\big(Q, \rho_j\big)\\
		&\hspace{2cm}+\big(\lambda_j-\rho_j\big) {R}_j -t {E}_{\mathrm{s}}\big(\lambda_j, P_V\big) \bigg]\bigg).
	\end{split}
\end{align}
Optimizing over the parameters gives
\begin{align}
	\begin{split}
		&p_e(\mcl{C}) \,\, \dot{\leq}\,\, \sum_{j=1}^{N_k} \inf_{\lambda_j>0} \, \inf_{\rho_j \geq 1} \exp\bigg(-n\bigg[\max_{Q \in \mcl{Q}_m} \, {E}_{\mathrm{x}}^{\prime}\big(Q, \rho_j\big)\\
		&\hspace{2cm}+\big(\lambda_j-\rho_j\big) {R}_j -t {E}_{\mathrm{s}}\big(\lambda_j, P_V\big) \bigg]\bigg)
	\end{split}\\
	\begin{split}
		&\hspace{0.9cm} \dot{\leq}\, \exp\Bigg(-n\bigg[\min_R \sup_{\lambda > 0} \sup_{\rho \geq 1} \bigg[ \max_{Q \in \mcl{Q}_m} \, {E}_{\mathrm{x}}^{\prime}\big({Q}, \rho\big)\bigg]\\
		&\hspace{2cm}+\big(\lambda-\rho\big)R-t {E}_{\mathrm{s}}\big(\lambda, P_V\big) \bigg]\Bigg).
	\end{split}
\end{align}
Thus, the resulting exponent is
\begin{align}
	\begin{split}
		&E_n(\mcl{C}_n) = \min_R \sup_{\lambda > 0} \sup_{\rho \geq 1} \bigg[ \max_{Q \in \mcl{Q}_m} \, {E}_{\mathrm{x}}^{\prime}\big({Q}, \rho\big)\bigg]\\
		&\hspace{2cm}+\big(\lambda-\rho\big) {R}-t {E}_{\mathrm{s}}\big(\lambda, P_V\big)
	\end{split}\\
	\begin{split}
		&\hspace{1cm}= \sup_{\lambda > 0}  \min_R \sup_{\rho \geq 1} \bigg[ \max_{Q \in \mcl{Q}_m} \, {E}_{\mathrm{x}}^{\prime}\big({Q}, \rho\big)\bigg]\\
		&\hspace{2cm}+\big(\lambda-\rho\big) {R}-t {E}_{\mathrm{s}}\big(\lambda, P_V\big), \label{eq-swap-rho-r}
	\end{split}
\end{align}
where we applied Sion’s minimax theorem \cite{1103040253} in \eqref{eq-swap-rho-r} to swap $\min_R$ and $\sup_{\lambda}$, since $E_s(\lambda, P_V)$ is convex in $\lambda$.
If $\lambda < 1$, then $(\lambda - \rho) < 0$, so $R \to \infty$, making the expression $-\infty$.
Therefore, $\lambda$ must satisfy $\lambda \geq 1$, leading to
\begin{align}
	\begin{split}
		E_n(\mcl{C}_n)
		&= \sup_{\lambda \geq  1}  \min_R \sup_{\rho \geq 1} \bigg[ \max_{Q \in \mcl{Q}_m} \, {E}_{\mathrm{x}}^{\prime}\big({Q}, \rho\big)\bigg]\\
		&\hspace{1cm}+\big(\lambda-\rho\big) {R}-t {E}_{\mathrm{s}}\big(\lambda, P_V\big),
	\end{split}
\end{align}
which completes the proof. Note that we write a minimum over $R$ rather than an infimum, since the optimal value is always achieved within the interval $0 \leq R \leq t H(V)$.
%This concludes the proof.

%%%%%%%%%%%%%%%%%%%%%%%%%%%%%%%%%%%%%%%%%%%%%%%%%%%%%%%%%%%%%%%%%%%%%%%%%%%%%%%%%%%%%%%%%%%%%%%%%%%%

\subsection{Proof of Theorem \ref{theorem-dual-concave-hull}}
% TODO:
Let $g(\lambda) \triangleq E_{\mathrm{x}}^{\prime}(\lambda, \mcl{Q}_m)$. 
Adopting from~\cite{1614076}, the concave conjugate of $g(\cdot)$ over $[1, \infty)$ is given by~\cite{rockafellar1970a}
\begin{align}
	g^{\star}(\lambda) = \inf_{\rho \geq 1} \Big\{ \lambda \rho - E_{\mathrm{x}}^{\prime}(\mcl{Q}_m, \rho)\Big\}.
\end{align}
The concave hull of $g(\cdot)$ is given by its biconjugate $g^{\star\star}(\lambda)$. Hence, we obtain
\begin{align}
	g^{\star\star}(\lambda) 
	&= \inf_{R} \bigg\{\lambda R - \inf_{\rho \geq 1} \Big\{  \rho R - E_{\mathrm{x}}^{\prime}(\mcl{Q}_m, \rho)\Big\}\bigg\}\\
	&= \inf_{R} \sup_{\rho \geq 1} \Big\{E_{\mathrm{x}}^{\prime}(\mcl{Q}_m, \rho) + (\lambda - \rho) R\Big\}.
\end{align}
This argument closely follows the proof of~\cite[Lemma~3]{1614076}.

%%%%%%%%%%%%%%%%%%%%%%%%%%%%%%%%%%%%%%%%%%%%%%%%%%%%%%%%%%%%%%%%%%%%%%%%%%%%%%%%%%%%%%%%%%%%%%%%%%%%

\subsection{Proof of Lemma \ref{lem-partition}}
Assume the two classes are associated with given parameters $\lambda_c$ and $\rho_c$. 
If $\bs{v} \in \mcl{T}^k(P_i)$ is assigned to class 1, then by~\eqref{eq-best-partition}, we have
\begin{align}
	\begin{split}
		&E_{\mathrm{x}}^{\prime}({Q}_1, \rho_1) + (\lambda_1 - \rho_1){R}_i-t {E}_{\mathrm{s}}(\lambda_1, P_V) \\
		&\hspace{0.75cm}\geq E_{\mathrm{x}}^{\prime}({Q}_2, \rho_2) + (\lambda_2 - \rho_2){R}_i-t {E}_{\mathrm{s}}(\lambda_2, P_V) .
	\end{split}
\end{align}
Rearranging terms, this implies
\begin{align}\label{eq-proof-partition}
	\!R_i \! \leq \! \frac{E_{\mathrm{x}}^{\prime}({Q}_1, \rho_1)\! -\!E_{\mathrm{x}}^{\prime}({Q}_2, \rho_2)\!  +\! t\left[ {E}_{\mathrm{s}}(\lambda_2, P_V)\!-\!t {E}_{\mathrm{s}}(\lambda_1, P_V)\right] }
	{\big(\lambda_2 - \rho_2 \big) - \big(\lambda_1 - \rho_1 \big)}.
\end{align}
Similarly, if $\bs{v} \in \mcl{T}^k(P_i)$ is assigned to class 2, the inequality in~\eqref{eq-proof-partition} is reversed.

%%%%%%%%%%%%%%%%%%%%%%%%%%%%%%%%%%%%%%%%%%%%%%%%%%%%%%%%%%%%%%%%%%%%%%%%%%%%%%%%%%%%%%%%%%%%%%%%%%%%

\subsection{Proof of Lemma \ref{lem-primal-nopart-opt}}

The proof follows the same steps as in Lemma~\ref{lem-dual-given}, except that all codewords now share the same composition. As a result, ${E}_{\mathrm{x}}(Q, \rho)$ appears in place of ${E}_{\mathrm{x}}^{\prime}(Q, \rho)$.

\end{document}